\documentclass[prd,floatfix,showpacs]{revtex4-1}
\usepackage{amsmath}
\usepackage{amsfonts}
\usepackage{amssymb}
\usepackage{graphicx}
\usepackage[margin=3.0cm]{geometry}
\usepackage{hyperref}

\newcommand\beq{\begin{equation}}      
\newcommand\beqnn{\begin{eqnarray*}}   
\newcommand\beqa{\begin{eqnarray}}     
\newcommand\beqann{\begin{eqnarray*}}  

\newcommand\eeq{\end{equation}}        
\newcommand\eeqnn{\end{eqnarray*}}     
\newcommand\eeqa{\end{eqnarray}}       
\newcommand\eeqann{\end{eqnarray*}}    

\newcommand\bi{\begin{itemize}}
\newcommand\ei{\end{itemize}}

\begin{document} 

\title{Spectral triplets, statistical mechanics and emergent geometry in non-commutative quantum mechanics}  
\date{\today}
\author{F. G. Scholtz$^a$ and B. Chakraborty$^{a,b}$}
\affiliation{$^a$Institute of Theoretical Physics, University of Stellenbosch, Stellenbosch 7600, South Africa\\
$^b$S.~N.~Bose National Centre for Basic Sciences,JD Block, Sector III, Salt Lake, Kolkata-700098, India}

\begin{abstract}

We show that when non-commutative quantum mechanics is formulated on the Hilbert space of Hilbert-Schmidt operators acting on a classical configuration space, spectral triplets as introduced by Connes in the context of non-commutative geometry arise naturally. A distance function as defined by Connes can therefore also be introduced.  We proceed to give a simple algorithm to compute this function in generic situations. Using this we compute the distance between pure and mixed states on quantum Hilbert space and demonstrate a tantalizing link between statistics and geometry.

\end{abstract}
\pacs{}

\maketitle
\section{Introduction}
The idea that space-time may be quantized in some sense and that a more abstract notion of geometry may be required has become firmly established in the physics literature in the past two decades.  Connes was the first to introduce, in a rigorous fashion, the notion of non-commutative geometry \cite{connes}, captured by a spectral triplet $({\cal A},{\cal H}, D)$ with ${\cal A}$ an involutive algebra acting on a Hilbert space ${\cal H}$ by a representation $\pi$ and $D$ is the so-called Dirac operator on ${\cal H}$.  The further technical conditions that this operator must satisfy can be found in \cite{connes}.   
 
In a separate, yet related development, it was argued by Doplicher et al \cite{Doplicher} from the considerations of both general relativity and quantum mechanics that the localization of an event in space-time with arbitrary accuracy is operationally impossible and this feature is captured by postulating a non-vanishing commutation relations between operator-valued coordinates. In its simplest form they  
are given as
\begin{eqnarray}
[\hat{x}_{\mu},\hat{x}_{\nu}]=i\theta_{\mu\nu},
\label{non}
\end{eqnarray}
where $\theta$ is anti-symmetric and its entries are viewed as new fundamental constants.  This form of non-commutativity also follows from the low energy limit of string theory \cite{Seiberg}.

Despite the fact that the two dimensional non-commutative (Moyal) plane also fits in the framework of non-commutative geometry as was established quite some time back for both compact \cite{a6} and non compact cases \cite{a7}, a detailed and explicit investigation into the link between these two versions of non-commutative space-time and associated geometries seems to have been started only recently with explicit computations of the Connes' spectral distance between pure state harmonic oscillator functions as well as translated states \cite{cag,mart}. One possible reason could be that the explicit computation of the Connes' distance function can be very challenging. Another reason may be that the Connes' approach was primarily applied to " almost commutative spaces " which are generically of the form $M \times F$ with $M$ representing $4-$dimensional ordinary commutative spacetime describing the gravitational part and $F$ corresponding to a zero$-$ dimensional discrete space described by matrices, which takes care of  the gauge part of the standard model \cite{a8}. The spaces of the type $M \times F$  has thus only a " mild " non-commutativity stemming from this internal space $F$. This is believed to be adequate to describe physics up to the GUT scale. However, at a still higher energy scale, like in the vicinity of the Planck scale, one may have to take the fuzziness of spacetime, captured by equations like (\ref{non}), seriously.   

Connes' spectral distance function relies on the notion of states, which are positive linear functionals of norm one, over the involutive algebra ${\cal A}$.  They are also closely related to the notion of density matrices, which describe the physical states of a quantum system.  This naturally raises the question whether the formulation of non-commutative quantum mechanics as set out in \cite{a10,a11}, which entails a representation of the quantum system on the Hilbert space of Hilbert-Schmidt operators, does not naturally encode geometric information in the context of Connes' non-commutative geometry.  This is the question we investigate here and we indeed find that this formulation of non-commutative quantum mechanics naturally encodes geometry in the form of spectral triplets.  Moreover we succeed in giving a simple algorithm to compute the Connes' spectral distance between states. It should, however, be remarked that the notion of spectral triplets that we introduce here is stronger than the conventional one, which takes the involutive algebra to be the space of bounded operators on $\cal H$, while here it is taken as the Hilbert-Schmidt operators and therefore is itself a Hilbert space.  It is precisely this Hilbert space structure of the involutive algebra that facilitates the present analysis.    

However, in contrast to the analysis in \cite{cag,mart} that used the Moyal star product, which, as shown in \cite{a9}, stems from a particular choice of basis in the quantum Hilbert space, our analysis is carried out in a basis independent operator approach that bypasses the use of any star product (as explained in \cite{a9}) and any ambiguities that may result therefrom.  Indeed, the analysis carried out here becomes problematic for the Moyal star product as the necessary positivity condition is not satisfied \cite{a9}.  On the other hand, the power of the operator approach was demonstrated by solving for the spectrum of a spherical well potential in a non-commutative plane \cite{a10} and the Coulomb problem in $3D$ non-commutative space \cite{a12} (with the coordinates satisfying a $SU(2)$ algebra ). The current approach and that of \cite{cag,mart} also differ on a more fundamental level in that the analysis in \cite{cag,mart} was essentially carried out in the classical configuration space $\mathcal{H}_c$ (see section \ref{one}). Here we generalize the notion of the Connes' distance to the true physical states of a non-commutative system that are density matrices on the quantum Hilbert space, $\mathcal{H}_q$, which has a natural tensorial structure $\mathcal{H}_q \equiv \mathcal{H}_c \otimes \mathcal{H}^*_c$ (see section \ref{one}). This offers rather tantalizing possibilities in terms of the modification of the implied geometry and the possible statistical underpinnings thereof.

The paper is organized as follows: In section \ref{one} we briefly review the formulation of non-commutative quantum mechanics on a two dimensional non-commutative (Moyal) plane and in section \ref{three} we provide a brief review of earlier notions of geometry in quantum mechanics, which used the inner product on the Hilbert space to induce a metric on the manifold of quantum states. In section \ref{two} we show how the results of \cite{cag} can be recovered from a spectral triplet defined on the classical configuration space and establish a link with the earlier notions of geometry discussed in section \ref{three}.  In section \ref{five} we proceed to introduce a spectral triplet on the quantum Hilbert space and introduce distances between mixed states. We also explore the possible statistical underpinnings of the emerging geometry.  Section \ref{four} summarizes and concludes the paper.

\section{Quantum mechanics on the non-commutative plane}
\label{one}

In two dimensions the non-commutative plane is defined through the commutation relations
\begin{equation}
[\hat{x_{i}},\hat{x_{j}}]= i\epsilon_{ij}\theta.
\end{equation}
where $\epsilon$ is the anti-symmetric tensor with $\epsilon_{12}=-\epsilon_{21}=1$.  We use a hat to emphasize the operator nature of these coordinates.  One constructs standard creation and annihilation operators $b^{\dagger}$ and $b$:
\begin{eqnarray}
\label{createann}
b=\frac{\hat{x_{1}}+i\hat{x_{2}}}{\sqrt{2\theta}}, b^{\dagger}=\frac{\hat{x_{1}}-i\hat{x_{2}}}{\sqrt{2\theta}}
\end{eqnarray}
and view the non-commutative plane as a boson Fock space spanned by the eigenstate $|n\rangle$ of the radial operator $b^{\dagger}b$. We refer to it as the classical configuration space $\mathcal H_{c}$:
\begin{equation}
\mathcal H_{c}= span\lbrace|n\rangle =\frac{1}{\sqrt{n!}}(b^{\dagger})^{n}|0\rangle \rbrace.
\end{equation}
This space plays the same role as the classical configuration space $\mathcal R^{2}$ in commutative quantum mechanics. Next we introduce the quantum Hilbert space in which the physical states of the system and the non-commutative Heisenberg algebra are to be represented. This is taken to be the set of Hilbert-Schmidt operators over $\mathcal H_{c}$ and we refer to it as the quantum Hilbert space,$\mathcal H_{q}$,
\begin{equation}
\mathcal H_{q}=\lbrace\psi:{\rm tr_{c}}(\psi^{\dagger}\psi)<\infty\rbrace,
\end{equation}
where the subscript c refers to tracing over $\mathcal H_{c}$. Elements of $\mathcal H_{q}$ are denoted by a round bracket $ |\psi)$ and the inner product is defined as
\begin{eqnarray}
\label{qhinner}
(\phi |\psi)={\rm tr_{c}}(\phi^{\dagger}\psi).
\end{eqnarray}        
We reserve $\dagger$ to denote hermitian conjugation on the classical Hilbert space, while $\ddagger$ denotes hermitian conjugation on the quantum Hilbert space.  Note that $\mathcal H_q$ has a natural tensor product structure in that it can be viewed as the tensor product of $\mathcal H_c$ and its dual, i.e., $\mathcal H_q=\mathcal H_c\otimes \mathcal H_c^\star$.  We can therefore also write the elements of $\mathcal H_q$ in the form $|\psi,\phi)\equiv||\psi\rangle\langle\phi|)$.  We shall use this notation quite often below and refer to $\psi$ as the left hand and $\phi$ as the right hand sector.

In units of $\hbar=1$ a unitary representation of the non-commutative Heisenberg algebra (we use a hat to emphasize the operator nature of the coordinates and momenta and to distinguish them from their classical counterparts)
\begin{eqnarray}
[\hat{x_{i}},\hat{x_{j}}]= i\epsilon_{ij}\theta,
\end{eqnarray}
\begin{eqnarray}
[\hat{x_{i}},\hat{p_{j}}]= i\delta_{ij},
\end{eqnarray}
\begin{eqnarray}
[\hat{p_{i}},\hat{p_{j}}]=0
\end{eqnarray}
is obtained by the following action:
\begin{eqnarray}
X_{i}|\psi)= |\hat{x_{i}}\psi),\\
P_{i}|\psi)=\frac{1}{\theta}|\epsilon_{ij}[\hat{x_{j}},\psi]).
\end{eqnarray}
Here we use capital letters to distinguish operators acting on the quantum Hilbert space from those acting on the classical Hilbert space.  For notational simplicity we also drop the hat notation as the capital letters distinguish these operators from their classical counterparts. It is also useful to introduce the following quantum operators
\begin{eqnarray}
\label{qop}
B&=&\frac{1}{\sqrt{2\theta}}\left(X_1+iX_2\right),\nonumber\\
B^\ddagger&=&\frac{1}{\sqrt{2\theta}}\left(X_1-iX_2\right),\nonumber\\
P&=&P_1 + iP_2,\nonumber\\
P^\ddagger &=& P_1 -i P_2.
\end{eqnarray}
These operators act as follow
\begin{eqnarray}
\label{aqo}
B|\psi) &=& |b\psi),\nonumber\\
B^\ddagger|\psi) &=& |b^\dagger\psi),\nonumber\\
P|\psi)&=& -i \sqrt{\frac{2}{\theta}}|[b,\psi]),\nonumber\\
P^\ddagger|\psi) &=& i\sqrt{\frac{2}{\theta}}|[ b^{\dagger},\psi]).
\end{eqnarray}

The interpretation of this quantum system now proceeds as for a standard one.  The only modification required is that position measurement must now be interpreted in the context of a weak measurement (Positive Operator Valued Measure) rather than a strong (Projective Valued Measurement). The essence of the construction is based on the minimal uncertainty states on non-commutative configuration space, which are the normalized coherent states \begin{equation}
\label{cs}
|z\rangle = e^{-z\bar{z}/2}e^{z b^{\dagger}} |0\rangle,
\end{equation}
where $z=\frac{1}{\sqrt{2\theta}}\left(x_1+ix_2\right)$ is a dimensionless complex number.  These states provide an overcomplete basis on the non-commutative configuration space.  Corresponding to these states one constructs a state in quantum Hilbert space as follows: Introduce the non-orthogonal projection operators $|z\rangle\langle z|$ on $\cal H_{\rm c}$ and define
\begin{equation}
\label{braz}
|z,z )\equiv||z\rangle\langle z|), \quad B|z,z)=z|z,z),
\end{equation}
which leads to the natural interpretation of $\left(x_1,x_2\right)$ as the dimensionful position coordinates.  These states provide an overcomplete set on the quantum Hilbert space in the form \cite{a11,chris}
\begin{equation}
\label{complstar}
{1}_q=\int \frac{dz d\bar{z}}{\pi} |z,z)e^{\stackrel{\leftarrow}{\partial_{\bar{z}}}\stackrel{\rightarrow}{\partial_z}}(z,z|,\quad e^{\stackrel{\leftarrow}{\partial_{\bar{z}}}\stackrel{\rightarrow}{\partial_z}}=\sum_{n=0}^\infty\frac{1}{n!}\stackrel{\leftarrow}{\partial_{\bar{z}}}^n\stackrel{\rightarrow}{\partial_z}^n
\end{equation}
where $\stackrel{\leftarrow}{\partial_{\bar{z}}}$ denotes differentiation to the left and $\stackrel{\rightarrow}{\partial_z}$ differentiation to the right.  The operators
\begin{equation}
\label{povm}
\pi_z=\frac{1}{2\pi\theta}|z,z) e^{\stackrel{\leftarrow}{\partial_{\bar{z}}}\stackrel{\rightarrow}{\partial_z}}(z,z|
\end{equation}
form a set of complete, positive, but non-orthogonal and unormalized projection operators \cite{chris}, i.e.,
\begin{equation}
\int d{\bar z} dz\;\pi_z=1_q\,\quad (\psi|\pi_z|\psi)\geq 0,\forall\psi\,,\quad \pi_z\pi_w\ne \delta(z-w)\,,\quad \pi_z^2\propto \pi_z.
\end{equation} 
They therefore provide a Positive Operator Valued Measure (POVM) that can be used to give a consistent probability interpretation by assigning the probability of finding the outcome of a position measurement to be $\left(x_1,x_2\right)$, given that the system is described by the density matrix $\rho$, to be
\begin{eqnarray}
P(x_1,x_2)={\rm tr_q}\left(\pi_z\rho\right),
\end{eqnarray}
where ${\rm tr_q}$ denotes the trace over quantum Hilbert space. In particular, for a pure state density matrix $\rho=|\psi ) (\psi|$, this probability is given by 
\begin{eqnarray}
P(x_1,x_2)={\rm tr_q}\left(\pi_z\rho\right)=\left(\psi|\pi_z|\psi\right)=\frac{1}{2\pi\theta}(\psi|z) e^{\stackrel{\leftarrow}{\partial_{\bar{z}}}\stackrel{\rightarrow}{\partial_z}}(z|\psi),
\end{eqnarray}
which differs from the commutative probability only by the presence of the Voros product \cite{a11}.

In \cite{chris,a13} it was shown that these operators can also be written as follows:
\begin{equation}
\label{povm1}
\pi_z=\frac{1}{2\pi\theta}\sum_n|z,n)(z,n|,
\end{equation}
where $|n\rangle$ is an arbitrary orthonormal basis in the classical configuration space, implying that this probability should be interpreted as the "total" probability of finding $\left(x_1,x_2\right)$ as the outcome of a position measurement, which is insensitive to the value of the additional quantum number $n$ labeling the right hand sector.

\section{Induced metric from the Hilbert space inner product}
\label{three}

It was realized some time ago that the inner-product on a generic Hilbert space $\cal H$ quite naturally induces a metric on manifolds of quantum states (unit rays) \cite{a14,ash}. In this section we briefly review this, however, instead of working with the quantum states as unit rays, which invariably restricts the formalism to pure states, we use the density matrix as the primary object that represents the states of a quantum system.  This has the advantage of generalizing this notion of geometry to mixed states and it also ensures that the results are manifestly gauge invariant. This will also establish a link to Connes' notion of spectral distance in the next section.

Let us therefore consider a $n$-parameter family of density matrices $\rho (\vec{s})$ with $\vec{s}=(s_1 ,...., s_n) \in \mathcal{R}^n$.  The density matrices have the usual properties of hermiticity $( \rho ^{\dagger} = \rho )$, unit norm $( {\rm tr}\rho (\vec{s}) = 1 ) $, semi-positivity and we assume that the map $s\mapsto \rho(s)$ is at least $C^1$. Realizing that the density matrices can be equipped with the natural inner product and trace norm on Hilbert-Schmidt operators,
\begin{eqnarray}
\label{A2}
(A,B)={\rm tr}(A^\dagger B),\quad \parallel A \parallel ^2 = (A,A), 
\end{eqnarray}
the distance between two neighbouring states (density matrices) with coordinates $\vec{s}$ and $(\vec{s} + d\vec{s})$ can be defined as
\begin{equation}
\label{A1}
d^2 (\vec{s} , \vec{s}+d\vec{s}) \equiv \frac{1}{2} \parallel \delta \rho \parallel ^2 = \frac{1}{2} (\partial _i \rho , \partial _j \rho) ds^i ds^j. 
\end{equation}

In the case of a family of pure state density matrices $\rho (\vec{s})=|\psi(\vec{s})\rangle\langle\psi(\vec{s})|$, $|\psi(\vec{s})\rangle\in {\cal H}$  it is a matter of straight forward verification that the metric tensor  
\begin{equation}
\label{A3}
g_{ij} \equiv \frac{1}{2} (\partial _i \rho , \partial _j \rho ),
\end{equation}
when re-written in terms of $\psi (\vec{s})$ and its derivative, yields \cite{a14,ash}
\begin{eqnarray}
\label{A4}
g_{ij} = Re(\partial _i \psi , \partial _j \psi) - A_i A_j.
\end{eqnarray}
Here $A_i = -i (\psi , \partial _i \psi)$ represents the pull-back of the connection 1-form of the $U(1)$ principal bundle onto the parameter space, and the exterior derivative of this connection 1-form $A=A_i ds^i$ gives the symplectic 2-form: $\sigma = dA = Im( \partial _i \psi , \partial _j \psi )ds^i\wedge ds^j$.  This can be further expressed in terms of the covariant derivative $D_i = (\partial _i  - i A_i) $ as 
$g_{ij} = (D_i\psi , D_j\psi)$.

Finally, $d^2 (\vec{s} , \vec{s}+d\vec{s})$ can indeed be shown to be the smallest distance in the space of rays in the projective Hilbert space \cite{a14}
\begin{equation}
\label{N}
d^2 (\vec{s} , \vec{s}+d\vec{s}) = D^2 (R_{\psi _1} , R_{\psi _1} ) = \inf_{\alpha_1,\alpha_2}\parallel \psi _1 e^{i\alpha _1} - \psi _2 e^{i\alpha _2} \parallel ^2  
\end{equation}
where $R_{\psi}$ represents the ray associated with the state $\psi $.

\section{Spectral triplets on classical configuration space}
\label{two}
We start our analysis of the geometrical content of the construction in section \ref{one} by identifying the spectral triplet $({\cal A},{\cal H}, D)$ where ${\cal A}={\cal H}_q$, ${\cal H}={\cal H}_c\otimes C^2$, the action of elements of $a\in {\cal A}$ on ${\cal H}$ is defined through the representation $\pi(a)(|\psi\rangle,|\phi\rangle)=(a|\psi\rangle,a|\phi\rangle)$ and the Dirac operator is defined as
\begin{eqnarray}
\label{diracop}
D=-i\sqrt{\frac{2}{\theta}}\left(\begin{array}{cc} 0&b^\dagger\\ b& 0\end{array}\right).
\end{eqnarray}
Here we have chosen the Dirac operator as in \cite{cag} to enable direct comparison with their results.  

States $\omega$ are positive linear functionals of norm one over ${\cal A}$. Pure states play a rather fundamental role and is defined as those functionals that cannot be written as a convex linear combination of two other functionals. The Connes' spectral distance between two states is then defined by
\begin{eqnarray}
\label{connesd}
d(\omega,\omega^\prime)&=&\sup_{a\in B}|\omega(a)-\omega^\prime(a)|,\nonumber\\
B&=&\{a\in {\cal A}: ||[D,\pi(a)]||_{\rm op}\le 1\},\nonumber\\
||A||_{\rm op}&=&\sup_{\phi\in {\cal H}}\frac{||A\phi||}{||\phi||}.
\end{eqnarray}
This is a rather difficult quantity to compute and therefore we first cast it into a more tractable form for the purpose of explicit computation and interpretation. For this we shall restrict our analysis to states satisfying the following conditions:
\begin{itemize}
\item The states $\omega$, $\omega^\prime$ are normal states (see ref \cite{a15} for the definition).
\item The states $\omega$ and $\omega^\prime$ are separately bounded on $B$, i.e., $\omega(a)<\infty$ and $\omega^\prime(a)<\infty$, $\forall a\in B$.
\item Let $V_0=\{a\in{\cal A}: ||[D,\pi(a)]||_{\rm op}=0\}$, then the states $\omega$, $\omega^\prime$ are such that  $\omega(a)-\omega^\prime(a)=0\,,\forall a\in V_0$.
\end{itemize}

The first two conditions are actually quite mild and essentially only implies that we restrict our analysis to states that can be represented by density matrices. Generally, the third condition places a restriction on the states for which our result for the Connes' distance holds, but under some rather generic conditions on the Dirac operator, that implies a certain irreducibility condition, it turns out that the result holds for all states satisfying the first two conditions.  To show this we note that their is a close link between normal states and density matrices.  Indeed, any normal state over ${\cal A}$ can be uniquely written as \cite{a15}
\begin{equation}
\omega(a)={\rm tr_c}(\rho_\omega a)  
\end{equation} 
where $\rho_\omega$ is a hermitian, semi-positive operator on ${\cal H}_c$ of trace one.  Note that since ${\rm tr_c}(\rho_\omega^2)\le {\rm tr_c}(\rho_\omega)=1$, $\rho_\omega$ is a Hilbert-Schmidt operator and therefore $\rho_\omega\in {\cal A}={\cal H}_q$.  If $\omega$ is a pure state, so is $\rho_\omega$, i.e. it is of the form $\rho_\omega=|\psi\rangle\langle\psi|$ for some $|\psi\rangle\in {\cal H}_c$. This result can actually be understood quite easily from the Hilbert space structure of ${\cal A}={\cal H}_q$ as it follows simply from the Riesz-Frechet theorem, where the properties of $\rho_\omega$ derive from the properties of $\omega$. 

Using this, and noting that our assumption $\omega(a)<\infty$ and $\omega^\prime(a)<\infty$, $\forall a\in B$ implies ${\rm tr_c}(\rho_\omega a) <\infty$ and ${\rm tr_c}(\rho_{\omega^\prime} a)<\infty$, $\forall a\in B$ and setting $d\rho=\rho-\rho^\prime$, we can write the third condition as 
\begin{itemize}
\item The states are such that 
\begin{equation}
\label{cond3}
{\rm tr_c}((d\rho) a)=(d\rho,a)=0\,,\forall a\in V_0.
\end{equation}
\end{itemize}
We use the notation $d\rho$ to indicate that we are interested in infinitesimal changes. Furthermore note that since ${\rm tr_c}(d\rho^2)\le 2(1-{\rm tr_c}(\rho\rho^\prime))\le 2$ from the semi-positiveness of $\rho$ and $\rho^\prime$, $d\rho$ is also a Hilbert-Schmidt operator, as one would also have expected from the Hilbert space structure of these operators. 

Let us consider a Dirac operator of the form (\ref{diracop}) 
\begin{eqnarray}
\label{diracop1}
D=-i\left(\begin{array}{cc} 0&c^\dagger\\ c & 0\end{array}\right),
\end{eqnarray}
with the requirement that $c$, $c^\dagger$ forms an irreducible pair on ${\cal H}_c$, i.e., $[c,a]=[c^\dagger,a]=0 \Rightarrow a\propto 1_c$. It is easily checked that $||[D,\pi(a)]||_{\rm op}=0\Rightarrow ||[c,a]||_{\rm op}=||[c^\dagger,a]||_{\rm op}=0 \Rightarrow [c,a]=[c^\dagger,a]=0 \Rightarrow a\propto 1_c$. Since the identity element is not Hilbert-Schmidt, the set $V_0$ contains only the null element, and the condition ${\rm tr_c}((d\rho) a)=0$ holds for all states satisfying the first two conditions. For the choice $c=b$ in (\ref{diracop}) it immediately follows that the Dirac operator  satisfies this condition since ${\cal H}_{\rm c}$ carries an irreducible representation of the oscillator algebra.  

Returning to the Connes' distance function, we can write
\begin{equation}
\label{temp}
d(\rho,\rho^\prime)=\sup_{a\in B}|{\rm tr_c}((d\rho )a)|=\sup_{a\in B}|(d\rho,a)|
\end{equation}

We are now in a position to determine the supremium on the right of (\ref{temp}).  Let us introduce the orthogonal complement, $V_0^\bot$, of $V_0$. Note from (\ref{cond3}) that $V_0$ is a subspace of the space of elements orthogonal to $d\rho$.  We can then decompose any $a\in{\cal A}$ uniquely into its component $a_0\in V_0$ and its component $a_1\in V_0^\bot$ (note that $[D,\pi(a)]\ne 0,\forall a\in V_0^\bot$) . From (\ref{cond3}) we can then write the Connes' distance function as
\begin{eqnarray}
\label{temp1}
d(\rho,\rho^\prime)&=&\sup_{a\in B^\prime}|{\rm tr_c}((d\rho )a)|=\sup_{a\in B^\prime}|(d\rho,a)|,\\
B^\prime&=&\{a\in V_0^\bot: ||[D,\pi(a)]||_{\rm op}\le 1\}\nonumber.
\end{eqnarray}

The first step is to find a lower bound.  To do this we note that the family of operators $\Lambda\equiv\{a\in{\cal A}: a=\lambda d\rho, 0<\lambda\le  \frac{1}{||[D,\pi(d\rho)]||_{\rm op}}\}\subset B^\prime$. Taking the extremal element $a= \frac{d\rho}{||[D,\pi(d\rho)]||_{\rm op}}$ yields the lower bound 
\begin{equation}
\label{ineq1}
d(\rho,\rho^\prime)\ge \frac{||d\rho||^2_{\rm tr}}{||[D,\pi(d\rho)]||_{\rm op}}.  
\end{equation}

Next we show that this is also an upper bound.  Consider any $a\in B^\prime$ and write it as $a=||a||_{\rm tr}\hat a$ where $||\hat a||_{\rm tr}=1$.  Then we have from (\ref{temp1})
\begin{equation}
||a||_{\rm tr}||[D,\pi(\hat a)]||_{\rm op}\le 1.
\end{equation}
Since $||[D,\pi(\hat a)]||_{\rm op}>0$, this places an upper bound $||a||_{\rm tr}\le \frac{1}{s}$ with $s=\inf_{\hat a\in B^\prime}||[D,\pi(\hat a)]||_{\rm op}$. We decompose $\hat a=\cos(\theta){\hat {d\rho}}+\sin(\theta){\hat {d\rho}}_\bot({\hat a})$ with ${\hat {d\rho}}$ and ${\hat {d\rho}}_\bot({\hat a})$ unit vectors parallel and orthogonal to $d\rho$. Note that both components are still in $V_0^\bot$ and that the latter depends on ${\hat a}$. Since $||[D,\pi(\hat a)]||_{\rm op}\le |\cos(\theta)|||[D,\pi({\hat {d\rho}})]||_{\rm op}+|\sin(\theta)|||[D,\pi({\hat {d\rho}}_\bot({\hat a}))]||_{\rm op}$, with the equality holding for $\theta=0,\frac{\pi}{2}$ we have $s\le s^\prime=\inf_{\hat a\in B^\prime}(|\cos(\theta)|||[D,\pi({\hat {d\rho}})]||_{\rm op}+|\sin(\theta)|||[D,\pi({\hat {d\rho}}_\bot({\hat a}))]||_{\rm op})$. Varying only the component of ${\hat a}$ along $d\rho$, i.e., varying $\theta$, while holding ${\hat {d\rho}}_\bot({\hat a})$ fixed (geometrically this corresponds to rotating ${\hat a}$ in the plane formed by ${\hat a}$ and $d\rho$), it is simple to see that the minimum is determined by the extremal points $\theta=0,\frac{\pi}{2}$, depending on whether $|[D,\pi(\hat {d\rho})]||_{\rm op}$ or $||[D,\pi({\hat {d\rho}}_\bot({\hat a}))]||_{\rm op}$ is the minimum (we can restrict ourselves without loss of generality to $\theta\in [0,\frac{\pi}{2}]$. Thus we conclude $s\le s^\prime=\min\{||[D,\pi({\hat d\rho})]||_{\rm op},\inf_{\hat a\in B^{\prime\prime}}(||[D,\pi({\hat a})]||_{\rm op})\}\equiv\min\{s_1,s_2\}$ with $B^{\prime\prime}=\{a\in B^\prime: (d\rho,a)=0\}$.  However, since for $\theta=0,\frac{\pi}{2}$ the equality holds in the triangular inequality, we must have $s=s^\prime$. Since the right hand side of (\ref{temp1}) vanishes for $a\in B^{\prime\prime}$, we conclude that regardless of whether $s_1$ or $s_2$ is the minimum, 
\begin{equation}
|(d\rho,a)|\le ||d\rho||_{\rm tr}||a||_{\rm tr}\le \frac{||d\rho||_{\rm tr}}{s_1}=\frac{||d\rho||^2_{\rm tr}}{||[D,\pi(d\rho)]||_{\rm op}}
\end{equation}
This establishes this value as an upper bound and therefore the supremium must satisfy      
\begin{equation}
\label{ineq2}
d(\rho,\rho^\prime)\le \frac{||d\rho||^2_{\rm tr}}{||[D,\pi(d\rho)]||_{\rm op}}.  
\end{equation}

Inequalities (\ref{ineq1}) and (\ref{ineq2}) together yield the desired result   
\begin{eqnarray}
\label{connesd1}
d(\rho,\rho^\prime)&=&\frac{{\rm tr_c}(d\rho)^2}{||[D,\pi(d\rho)]||_{\rm op}}.
\end{eqnarray}
This is a much more tractable form for Connes' spectral distance, which clearly has a strong resemblance with (\ref{A1}) in the previous section, although the computation of the operator norm can still be tedious.  In fact, in the examples we compute below it turns out that if we replace the operator norm with the trace norm it only modifies the distance function by a trivial numerical constant.  Although a general proof that this is always the case lacks, it does suggest that defining a distance function, closely related to the Connes' distance function, by replacing the operator norm with the trace norm in (\ref{connesd1}) is sensible and leads to simplification since the trace norm is often easier to compute. 

It is worthwhile highlighting the role of (\ref{cond3}) in deriving (\ref{connesd1}). If this condition does not hold, we can find a $a_0\in V_0$ such that $(d\rho,a_0)\ne 0$.  Since (\ref{connesd}) now places no constraint on $||a_0||_{\rm tr}$, it is clear that no upper bound can be derived for $(d\rho,a_0)$ and it is in principle unbounded.  


Let us apply this to the example in \cite{cag} where the distance between harmonic oscillator states is calculated.  In the harmonic oscillator basis we can write \cite{note1} $a=\sum_{m,n} a_{m,n}|m\rangle\langle n|$. 
The states are taken as $\omega_m(a)=a_{m,m}$ with corresponding pure state density matrix $\rho_m=|m\rangle\langle m|$. We then take $d\rho=\rho_{m+1}-\rho_m$.  The calculation is now straightforward:
\begin{eqnarray}
\label{diraqsq}
\left[D,\pi(d\rho)\right]&=&-i\sqrt{\frac{2}{\theta}}\left(
\begin{array}{cc} 0&\left[b^\dagger,d\rho\right]\\ \left[b,d\rho\right]& 0\end{array}\right),\nonumber\\
\left[D,\pi(d\rho)\right]^\dagger\left[D,\pi(d\rho)\right]&=&\frac{2}{\theta}\left(\begin{array}{cc} \left[b,d\rho\right]^\dagger\left[b,d\rho\right]&0\\ 0& \left[b^\dagger,d\rho\right]^\dagger\left[b^\dagger,d\rho\right]\end{array}\right).
\end{eqnarray}

A simple calculation yields 
\begin{eqnarray}
\label{drhosqho}
\left[b,d\rho\right]^\dagger\left[b,d\rho\right]&=&4(m+1)|m+1\rangle\langle m+1|+m|m\rangle\langle m|+(m+2)|m+2\rangle\langle m+2|,\nonumber\\
\left[b^\dagger,d\rho\right]^\dagger\left[b^\dagger,d\rho\right]&=&(m+2)|m+1\rangle\langle m+1|+4(m+1)|m\rangle\langle m|+m|m-1\rangle\langle m-1|.
\end{eqnarray}
As both of these operators are already diagonal, the operator norm, which is nothing but the largest eigenvalue can be read off exactly. This yields 
\begin{equation}
||[D,\pi(d\rho)]||_{\rm op}=2\sqrt{\frac{2(m+1)}{\theta}}
\end{equation}
and  
\begin{equation}
\label{hodist}
d(m+1,m)=\sqrt{\frac{\theta}{2(m+1)}},
\end{equation}
which agrees precisely with the result of \cite{cag}. 

Next we calculate the distance between two coherent states.  We define the usual harmonic oscillator coherent states in ${\cal H}_c$ given by $|z>=e^{-\frac{{\bar z} z}{2}}e^{z b^\dagger}|0\rangle$ and consider the corresponding pure state density matrix $\rho=|z\rangle\langle z|$.  Then
\begin{eqnarray}
\label{B1}
d\rho &=& |z+dz\rangle\langle z+dz|-|z\rangle\langle z|\nonumber\\
&=&\left(b^{\dagger} - \bar{z}\right)|z\rangle \langle z| dz + |z\rangle \langle z| \left( b - z  \right)d\bar{z}+O(dz^2,d\bar z^2).
\end{eqnarray}
This can be written more simply as
\begin{equation}
d\rho=d\bar z|\tilde 0\rangle\langle\tilde 1|+dz |\tilde 1\rangle\langle \tilde 0| 
\end{equation}
where we have introduced new bosons $\tilde b=b-z$, $\tilde b^\dagger=b^\dagger-\bar z$ and the vacuum $|\tilde{0}\rangle=|z\rangle$ for which $\tilde b|\tilde{0}\rangle=0$, $|\tilde 1\rangle=\tilde b^\dagger|\tilde 0\rangle$. From this 
\begin{equation}
[b,d\rho]=dz|\tilde{0}\rangle\langle \tilde{0}|-\sqrt{2}d\bar z|\tilde{0}\rangle\langle \tilde{2}|-dz |\tilde{1}\rangle\langle \tilde{1}| 
\end{equation}
and 
\begin{eqnarray}
\label{drhosqcs}
\left[b,d\rho\right]^\dagger\left[b,d\rho\right]&=&d\bar zdz|\tilde{0}\rangle\langle \tilde{0}|-\sqrt{2}(d\bar z)^2|\tilde{0}\rangle\langle \tilde{2}|-\sqrt{2}(dz)^2|\tilde{2}\rangle\langle \tilde{0}|+2d\bar z dz|\tilde{2}\rangle\langle \tilde{2}|+d\bar zdz |\tilde{1}\rangle\langle \tilde{1}|,\nonumber\\ 
\left[b^\dagger,d\rho\right]^\dagger\left[b^\dagger,d\rho\right]&=&3d\bar zdz|\tilde{0}\rangle\langle \tilde{0}|+d\bar zdz |\tilde{1}\rangle\langle \tilde{1}|. 
\end{eqnarray}
This yields, on diagonalising the first one, 
\begin{equation}
||[D,\pi(d\rho)]||_{\rm op}=\sqrt{\frac{6 d{\bar z}dz}{\theta}}
\end{equation} 
and
\begin{equation}
\label{csdist}
d(z+dz,z)=\sqrt{\frac{2\theta}{3}}\sqrt{d{\bar z}dz},
\end{equation}
yielding the Euclidean distance.  One can quickly check from (\ref{drhosqho}) and (\ref{drhosqcs}) that replacing the operator norm in (\ref{hodist}) and (\ref{csdist}) by the trace norm one recovers, up to numerical factors, the same results.  In what follows we shall, for computational simplicity, therefore often replace the operator norm with the trace norm and rather compute the closely related distance function
\begin{eqnarray}
\label{connesd2}
\tilde d(\rho,\rho^\prime)&=&\frac{{\rm tr_c}(d\rho)^2}{||[D,\pi(d\rho)]||_{\rm tr}}.
\end{eqnarray}
In the examples above this distance function and the Connes' distance function only differ by a numerical constant. Whether this holds generally is an open question that needs further investigation.

\section{Spectral triplets on quantum Hilbert space}
\label{five}
In the construction of the previous section there is a one-to-one correspondence between states (points) $|\psi\rangle$ in the classical configuration space and pure state density matrices $\rho=|\psi\rangle\langle\psi|$ acting on classical configuration space. These density matrices are also elements of the quantum Hilbert space corresponding to diagonal states $|\psi,\psi)\equiv||\psi\rangle\langle\psi|)$.  However, there are many more physical pure states in the quantum Hilbert space of off-diagonal form $|\psi,\phi)\equiv||\psi\rangle\langle\phi|)$.  These states are in a one-to-one relation with the pure state density matrices $\rho_q=|\psi,\phi)(\psi,\phi|$ acting on quantum Hilbert space and in a many-to-one relation with states (points) in classical configuration space, while only the diagonal density matrices $\rho_q=|\psi,\psi)(\psi,\psi|$ are in a one-to-one correspondence with states in the classical configuration space. Note that this is quite different to commutative quantum mechanics where there is a one-to-one relation between pure states density matrices and points in configuration space. Specifically, in commutative quantum mechanics a particle localized at a point $x$ is described by the pure state density matrix $\rho=|x\rangle\langle x|$.  On the other hand, as discussed in section \ref{one} and further below, in non-commutative quantum mechanics a particle maximally localized at a point $(x_1,x_2)$, without any prior knowledge of the right hand sector, is described by the mixed state density matrix $\rho_q=\sum_n|z,n)(z,n|$ ($z=x_1+ix_2$).  When the right hand sector is also specified the particle is described by a pure state density matrix $\rho=|z,n)(z,n|$, but note that there are infinitely many choices for the right hand sector given a point $(x_1,x_2)$ in classical configuration space.   

We would therefore also like to calculate distances between off-diagonal pure states and mixed states on quantum Hilbert space.  To do this we introduce a further spectral triplet $({\cal A},{\cal H}, D)$ on quantum Hilbert space where ${\cal A}$ are the Hilbert-Schmidt operators on $\mathcal H_q$, ${\mathcal H}={\mathcal H}_q\otimes C^2$, the action of elements of $a\in {\cal A}$ on ${\cal H}$ is defined through the representation $\pi(a)(|\psi),|\phi))=(a|\psi),a|\phi))$ and the Dirac operator is defined as 
\begin{eqnarray}
\label{diracq}
D=-i\sqrt{\frac{2}{\theta}}\left(\begin{array}{cc} 0&B^\ddagger\\ B& 0\end{array}\right).
\end{eqnarray}   
Note that the Dirac operator defined here acts only on the left sector as it does not involve any momentum operators. 

Clearly, under the same conditions as in section \ref{two} on the states, the analysis that led up to (\ref{connesd1}) holds here with the replacement ${\cal H}_c\rightarrow {\cal H}_q$, i.e., 
\begin{eqnarray}
\label{connesdq}
d(\rho_q,\rho_q^\prime)&=&\frac{{\rm tr_q}(d\rho_q)^2}{||[D,\pi(d\rho_q)]||_{\rm op}}.
\end{eqnarray}
However, since the Dirac operator (\ref{diracq}) only acts on the left sector, the condition (\ref{cond3}) becomes non-trivial. Indeed it is simple to verify that the general operators that commute with the Dirac operator are of the form 
\begin{equation}
\Gamma=\sum_p\sum_{k,\ell}a_{k,\ell}|p,k)(p,\ell|
\end{equation}    
with $|p\rangle$ an orthonormal basis in classical configuration space and $a_{k,\ell}$ arbitrary complex numbers independent from $p$. In this case these operators span $V_0$. Generically $(d\rho_q,\Gamma)\ne 0$, (\ref{cond3}) is violated and we cannot expect (\ref{connesdq}) to yield the Connes' distance between all states. At best it will yield a lower bound, while the true Connes' distance may even be unbounded.    However, (\ref{connesdq}) still applies for states such that $(d\rho_q,\Gamma)=0,\;\forall\Gamma$ and can be used to compute the Connes' distance function between such states. Below we apply (\ref{connesdq}) in this context.  Indeed, we shall compute (\ref{connesdq}) regardless of whether the condition $(d\rho_q,\Gamma)=0$ is met or not and simply keep in mind that this will only yield the true Connes' distance when this condition holds. This is particularly useful in the context of a variational calculation as performed below.
 
For an infinitesimal change $d\rho_q$ in the density matrix $\rho_q$, a calculation paralleling the calculation leading to equation (\ref{diraqsq}) yields
\begin{eqnarray}
\label{diracsqq}
\left[D,\pi(d\rho_q)\right]&=&-i\sqrt{\frac{2}{\theta}}\left(
\begin{array}{cc} 0&\left[B^\ddagger,d\rho_q\right]\\ \left[B,d\rho_q\right]& 0\end{array}\right),\nonumber\\
\left[D,\pi(d\rho_q)\right]^\ddagger\left[D,\pi(d\rho_q)\right]&=&\frac{2}{\theta}\left(\begin{array}{cc} \left[B,d\rho_q\right]^\ddagger\left[B,d\rho_q\right]&0\\ 0& \left[B^\ddagger,d\rho_q\right]^\ddagger\left[B^\ddagger,d\rho_q\right]\end{array}\right).
\end{eqnarray}

To start we compute the distance between the pure states corresponding to the density matrices $\rho_q(m,\phi)=|m,\phi)(m,\phi|$ with $|m\rangle$ a harmonic oscillator state and $|\phi\rangle$ an arbitrary state in the classical configuration space.  Introducing $d\rho_q=|m+1,\phi)(m+1,\phi|-|m,\phi)(m,\phi|$, the calculation proceeds exactly as in section \ref{two}.  Indeed, the state $|\phi\rangle$ is purely a spectator and the result obtained is exactly the same as in (\ref{hodist}).  The same is true for the pure states described by the density matrix $\rho_q(z,\phi)=|z,\phi)(z,\phi|$ where $|z\rangle$ is a coherent state in the classical configuration space.  Computing the distance between these states yields again the result (\ref{csdist}), independent of $|\phi\rangle$.  The independence of these results from the choice of the right hand sector specified by $|\phi\rangle$ is of course a consequence of the choice of the Dirac operator, which only acts on the left hand sector.  In these cases it is also simple to verify that $(d\rho_q,\Gamma)=0,\;\forall\Gamma$ and thus this will yield the true Connes' distance function.

We can use the same Dirac operator to compute the more general spectral distance between the states $\rho _q(m,\phi) = |m,\phi )(m,\phi|$ and $\rho_q(m+1,\phi^\prime)= |m+1,\phi ^{\prime} )(m+1,\phi ^{\prime}|$ where $\phi ^{\prime}$ is different and orthogonal to $\phi$, i.e., $\langle\phi|\phi^\prime\rangle=\delta_{\phi,\phi^\prime}$. Introducing $d\rho _q = | m+1,\phi ^{\prime} )  ( m+1,\phi ^{\prime}| - |m ,\phi)  ( m,\phi| $, we get
\begin{eqnarray}
\label{b2}
[B, d\rho _q]^\ddagger [B, d\rho _q] &=& (m+1)\mid m+1, \phi ^{\prime} ) ( m+1, \phi ^{\prime} \mid + (m+1) \delta _{\phi \phi{\prime}}\mid m+1, \phi ^{\prime} ) ( m+1, \phi \mid + m \mid m, \phi ) ( m, \phi \mid \nonumber \\
&+& (m+2)\mid m+2, \phi ^{\prime} ) ( m+2, \phi ^{\prime} \mid + (m+1) \delta _{\phi \phi{\prime}}\mid m+1, \phi ^{\prime} )( m+1, \phi ^{\prime} \mid \nonumber \\
&+& (m+1)\mid m+1, \phi )( m+1, \phi \mid,
\end{eqnarray}
\begin{eqnarray}
\label{b3}
[B^{\ddagger}, d\rho _q]^\ddagger [B^{\ddagger}, d\rho _q] &=& (m+2)\mid m+1, \phi ^{\prime} ) ( m+1, \phi ^{\prime} \mid + (m+1) \mid m, \phi )( m, \phi \mid  + (m+1) \delta _{\phi \phi{\prime}}\mid m, \phi ) ( m, \phi ^{\prime} \mid \nonumber \\ 
&+&(m+1) \delta _{\phi \phi{\prime}}\mid m, \phi ^{\prime} ) ( m, \phi ^{\prime} \mid + (m+1)\mid m, \phi ^{\prime} ) ( m, \phi ^{\prime} \mid + m \mid m-1, \phi ) ( m-1, \phi \mid.\\ 
\end{eqnarray}
It is easily seen that for $\phi = \phi ^{\prime} $ we recover the earlier result (\ref{hodist}), but for $\phi \neq \phi ^{\prime}$ we get 
\begin{eqnarray}
\label{B4}
||[D,\pi(d\rho_q)]||_{\rm op}=\sqrt{\frac{2(m+2)}{\theta}}
\end{eqnarray}
This eventually yields $d(\rho(m+1, \phi ^{\prime}) , \rho(m, \phi)) = \sqrt{\dfrac{2\theta}{m+2}}$. In this case, however, $(d\rho_q,\Gamma)\ne 0,\;\forall\Gamma$ so that this does not correspond to the true Connes' distance.  However, taking this quantity at face value, it does show that $d(\rho(m+1,\phi), \rho(m,\phi)) < d(\rho(m+1, \phi ^{\prime}) , \rho(m, \phi)) $ and demonstrates that, despite the fact that the Dirac operator only acts on the left hand sector, the distance between two states depends on the right hand sector and, indeed, that the distance increases when the right hand sectors are taken differently.  Here we have taken the right hand to be a pure state that depends on the left hand, i.e., it changes from point to point. One may contemplate an even more general situation by taking the right hand sector to be a statistical mixture that changes from point to point, i.e., we can consider the following density matrices
\begin{eqnarray}
\label{densmat}
\rho_q(m)&=&\sum_n p_n(m)|m,n)(m,n|,\quad \sum_n p_n(m)=1,\forall m,\nonumber\\  
\rho_q(z,\bar z)&=&\sum_n p_n(z,\bar z)|z,n)(z,n|,\quad \sum_n p_n(z,\bar z)=1,\forall z,  
\end{eqnarray}
where $|m\rangle$ and $|z\rangle$ are as above, while $|n\rangle$ is an arbitrary orthonormal basis in classical configuration space and $p_n$ a set of position dependent probabilities.  To clarify the physical meaning of these states, we compute the average of the radial operator in $\rho _{q}(m)$, which simply yields ${\rm tr}_q (B_L^\ddagger B_L\rho_q(m))=m$.  As the fluctuations vanish, these are states localized at a fixed radial distance $m$. Similarly the averages of $(B^\ddagger+B)/2$ and $i(B^\ddagger-B)/2$ in the state $\rho_q(z,\bar z)$ yield the real and imaginary parts of $z$, while the their fluctuations saturate the minimal uncertainty relation.  This implies that these states describe a system maximally localized at the point $\left(x_1,x_2\right)$. Note that the operators $\pi_z$ introduced in (\ref{povm1}) are density matrices of this type as they can be interpreted as the density matrix for a particle maximally localized at $z$, but with complete ignorance of the right hand sector, i.e., all probabilities are equal. Indeed, computing the probability of finding a particle at point $z^0=(x^0_1+ix^0_2)/\sqrt{2\theta}$, given that it is prepared in the state described by the density matrix $\pi_z$ with $z=(x_1+ix_2)/\sqrt{2\theta}$, yields a Gaussian $P(x_1,x_2)\propto e^{-\frac{1}{2\theta}((x_1-x^0_1)^2+(x_2-x^0_2)^2)}$.  It is also for this reason that their use as a POVM leads to a measurement of position that disregards any information of the right hand sector.

In these cases it is easily verified that if the probabilities $p_n$ are position independent, i.e., they do not depend on $m$ or $z$, $(d\rho_q,\Gamma)=0,\;\forall\Gamma$.  Thus in this case (\ref{connesdq}) will yield the true Connes' distance between these states.

The choice of probabilities in (\ref{densmat}) is, of course, an open issue.  In equilibrium statistical mechanics, the choice of probabilities is dictated by equilibrium considerations, i.e., one maximizes the entropy subject to constraints on averages, e.g., the average energy that leads to a Boltzman distribution. Here it is not clear which criteria should be used to determine the probability distribution, but two obvious possibilities present themselves.  The first is to choose the probabilities such that the path length between two points is minimized.  The second is to fix the probabilities from a condition of local thermal equilibrium, i.e., the local average energy is fixed and the local entropy associated with the density matrices (\ref{densmat}) is optimized. The rest of this section explores the interplay between geometry and statistics from these two perspectives.  Ideally one would prefer that the probability distributions and geometries that emerge from these two perspectives coincide.  However, it turns out below that this only happens in the $T\rightarrow\infty$ limit.  

To facilitate the computation below, we use the modified distance function (\ref{connesd2}), adapted to quantum Hilbert space, rather than (\ref{connesdq}), i.e., 
\begin{eqnarray}
\label{connesdq1}
\tilde d(\rho_q,\rho_q^\prime)&=&\frac{{\rm tr_q}(d\rho_q)^2}{||[D,\pi(d\rho_q)]||_{\rm tr}}.
\end{eqnarray}
Setting $d\rho_q=\rho_q(m+1)-\rho_q(m)$ one obtains
\begin{equation}
\label{distmix1}
\tilde d(m+1,m)=\frac{\sqrt{\theta}}{2}\frac{\sum_n\left(p_n^2(m+1)+p_n^2(m)\right)}{\sqrt{\sum_n\left((2m+3)p_n^2(m+1)+(2m+1)p_n^2(m)+2(m+1)p_n(m+1)p_n(m)\right)}}.
\end{equation}

Let us first consider the first perspective and ask which choice of probabilities actually minimizes the distance function.  Starting with (\ref{distmix1}), the distance between two points $m_i$ and $m_f$ is given by 
\begin{equation}
\tilde d(m_f,m_i)=\sum_{m=m_i}^{m_f-1}\tilde d(m+1,m).
\end{equation}
One easily concludes that the probabilities that minimize the distance must satisfy the equation
\begin{equation}
\label{min}
\Delta p_n=\lambda,\;\forall n
\end{equation}  
where
\begin{equation}
p_n =\left(\begin{array}{cc} p_n(m_i)\\ \vdots\\ p_n(m_{i+1})\\ p_n(m_f)\end{array}\right),\quad \lambda=\left(\begin{array}{cc} \lambda(m_i)\\ \vdots\\ \lambda(m_{i+1}) \\ \lambda(m_f)\end{array}\right),\nonumber\\
\end{equation}
$\lambda(m)$ are the Lagrange multipliers imposing the constraints that the probabilities sum to one and $\Delta$ is the tri-diagonal matrix 
\begin{equation}
\Delta = \left(\begin{array}{ccccccc} a(m_i)&b(m_i)\\ b(m_i)&a(m_{i+1})&b(m_{i+1})\\\ddots&\ddots&\ddots\\ &b(m_f)&a(m_f)\nonumber\\
\end{array}\right).
\end{equation}
The matrix elements of the matrix $\Delta$ are given by
\begin{eqnarray}
a(m)&=&\sqrt{\theta}\left(g(m)+g(m-1)+(2m+1)(f(m)+f(m-1))\right),\nonumber\\
b(m)&=&-\sqrt{\theta}(m+1)f(m+1),\nonumber\\
g(m)&=&\frac{1}{\sqrt{\sum_n\left((2m+3)p_n^2(m+1)+(2m+1)p_n^2(m)+2(m+1)p_n(m+1)p_n(m)\right)}},\nonumber\\
f(m)&=&\frac{1}{2}\frac{\sum_n\left(p_n^2(m+1)+p_n^2(m)\right)}{\left(\sum_n\left((2m+3)p_n^2(m+1)+(2m+1)p_n^2(m)+2(m+1)p_n(m+1)p_n(m)\right)\right)^{3/2}}.
\end{eqnarray}
It is important to note that although $\Delta$ is a function of all the probabilities, it is independent of $n$. Using this we can sum over $n$ in (\ref{min}), which shows that $\lambda=\Delta p$, where $p$ is the column vector with equal entries $1/\Omega$, $\Omega$ being a cut-off for the sum over $n$. Substituting this back in (\ref{min}) shows that the probabilities are independent from $m$ and $n$. In this case the distance function reduces to    
\begin{equation}
\label{dist1}
\tilde d(m+1,m)=\frac{\sqrt{\theta}}{\sqrt{6\Omega}}\frac{1}{\sqrt{m+1}},
\end{equation}
which only differs from (\ref{hodist}) by a probability dependent global scale factor. In this case $(d\rho_q,\Gamma)=0,\;\forall\Gamma$ and (\ref{dist1}) should be closely related to the Connes' distance, the only difference residing in the use of the trace norm in (\ref{connesdq1}), rather than the operator norm.  We expect this to only yield numerical factors.

Next we explore the geometry emerging from the second perspective.  To do this we introduce a local entropy associated with the density matrices (\ref{densmat}) in the usual manner
\begin{eqnarray}
\label{ent}
S(m)&=&-\sum_n p_n(m)\log p_n(m),\nonumber\\  
S(z,\bar z)&=&-\sum_n p_n(z,\bar z)\log p_n(z,\bar z).
\end{eqnarray}
A simple calculation shows that maximizing this local entropy leads, as above, to equally distributed probabilities.  This, of course, corresponds to the $T\rightarrow\infty$ limit where only entropy plays a role and all states are assigned equal probabilities.  
  
If we associate an energy scale $\epsilon_n$ with the right hand states labeled by $n$ and require the local average energy to be fixed on $E(m)$, maximization of the local entropy yields a local Boltzman distribution
\begin{equation}
p_n(m)=\frac{e^{-\beta(m)\epsilon_n}}{Z(\beta(m))},\quad Z(\beta(m))=\sum_n e^{-\beta(m)\epsilon_n}, 
\end{equation}
where $\beta(m)$ is the local inverse temperature.  From this the distance function (\ref{distmix1}) can be computed.  In particular, if we assume the local average energy and hence the temperature to be independent of $m$, we obtain
\begin{equation}
\tilde d(m+1,m)=\frac{\sqrt{\theta}}{\sqrt{6}}\frac{\sqrt{Z(2\beta)}}{Z(\beta)}\frac{1}{\sqrt{m+1}},
\end{equation}
which again only differs from (\ref{hodist}) by a temperature dependent global scale factor. The same remarks as for (\ref{dist1}) regarding the relation to the Connes' distance are applicable here.  

This implied modified geometry for mixed states becomes even more explicit in the case of the continuous coherent state basis, which we consider next.

The distance between the coherent state density matrices simplifies considerably with a slight redefinition of the Dirac operator.  We take the Dirac operator to be 
\begin{eqnarray}
\label{distmix2}
D&=&-i\sqrt{\frac{2}{\theta}}\left(\begin{array}{cc} 0& \left[B^\ddagger,2\pi\theta\pi_z\right]\\ \left[B,2\pi\theta\pi_z\right]& 0\end{array}\right)\\
&=&-i\sqrt{\frac{2}{\theta}}\left(\begin{array}{cc} 0&\sum_n|\tilde 1,n)(\tilde 0,n|\\ \sum_n|\tilde 0,n)(\tilde 1,n|& 0\end{array}\right).
\end{eqnarray}   
Here we have again introduced the bosons $\tilde b=b-z$, $\tilde b^\dagger=b^\dagger-\bar z$ and the vacuum $|\tilde{0}\rangle=|z\rangle$.  Note that the Dirac operator is still insensitive to the right hand sector.  Apart from the simplification that occurs with this choice of the Dirac operator, it is also a more natural choice in that it is based on the local vacuum $|\tilde{0}\rangle=|z\rangle$ and local excitations $\tilde b^\dagger=b^\dagger-\bar z$ around this vacuum. The same remarks as made for the Dirac operator (\ref{diracq}) apply to this Dirac operator.  The general operators that commute with it are of the form 
\begin{equation}
\tilde\Gamma=\sum_p\sum_{k,\ell}a_{k,\ell}|\tilde p,k)(\tilde p,\ell|
\end{equation}    
As before, it is also straightforward to verify that for the density matrices (\ref{densmat}), $(d\rho_q,\tilde\Gamma)=0,\;\forall\tilde\Gamma$ if the probabilities are position independent.

The computation now follows the same route as before and one obtains
\begin{equation}
\tilde d(z+dz,z)=\frac{\sqrt\theta}{2}\sqrt{2\sum_n\left(p_n(z)^2+\frac{\partial p_n(z)}{\partial \bar{z}}\frac{\partial p_n(z)}{\partial z}\right)d\bar{z}dz+\sum_n\left(\left(\frac{\partial p_n(z)}{\partial \bar{z}}\right)^2d\bar{z}^2+\left(\frac{\partial p_n(z)}{\partial z}\right)^2dz^2\right)}.
\end{equation} 

This clearly exhibits a modified geometry.  Assuming that $p_n$ is just a function of the dimensionful radial coordinate $r=\sqrt{\frac{\bar{z}z\theta}{2}}$ one obtains
\begin{equation}
\tilde d(z+dz,z)=\frac{1}{2\sqrt{2}}\sqrt{\sum_n\left(2p_n(r)^2+{p^\prime_n(r)}^2\right)}\sqrt{dr^2+r^2d\phi^2},
\end{equation} 
which is related to the Euclidean distance by a probability dependent conformal transformation.  Note that since the probabilities are position dependent, this will not correspond to the true Connes' distance. We can of course again compute the probabilities from the condition of local equilibrium.  In the $T\rightarrow\infty$ limit, the probabilities are obtained by maximizing (\ref{ent}), which yields equal, $r$-independent probabilities and subsequently, up to a global scale, an Euclidean geometry, which should be closely related to Connes' distance function. If we constrain the local average energy to $E(r)$, the temperature becomes position dependent and the probabilities again follow a Boltzman distribution
\begin{equation}
p_n(r)=\frac{e^{-\beta(r)\epsilon_n}}{Z(\beta(r))},
\end{equation}
yielding a conformal scale factor. If the local average energy is position independent, so is the temperature, and we obtain an Euclidean geometry with a global, temperature dependent, scale factor
\begin{equation}
\tilde d(z+dz,z)=\frac{\sqrt{Z(2\beta)}}{2Z(\beta)}\sqrt{dr^2+r^2d\phi^2}.
\end{equation} 
As the prefactor is bounded by $0<\frac{\sqrt{Z(2\beta)}}{2Z(\beta)}\le \frac{1}{2}$, reaching its minimum at $T\rightarrow\infty$ and its maximum at $T\rightarrow 0$ this suggest, quite remarkably, that distances expand as the global temperature is lowered.  We expect this distance function to be very closely related to the true Connes' distance as the probabilities are position independent and therefore $(d\rho_q,\tilde\Gamma)=0,\;\forall\tilde\Gamma$.  The only difference from the true Connes' distance is therefore the use of the trace, rather than operator norm, which we expect to yield only numerical prefactors.

\section{Conclusions}
\label{four}

The formulation of non-commutative quantum mechanics on the Hilbert space of Hilbert-Schmidt operators naturally encodes the notion of spectral triplets and thus geometry, albeit in a sense stronger than in the conventional sense.  Conditions under which a more tractable form of the Connes' distance formula apply have been derived.  This form, and natural modifications of it, were used to make the geometry explicit on the level of the classical configuration space and the quantum Hilbert space.  In the former case, as in commutative quantum mechanics, there is a one-to-one correspondence between pure state density matrices and points in the classical configuration space.  In the latter case this no longer holds due to the tensor product structure of the quantum Hilbert space and the distance function depends on information encoded in the right hand sector, even though the Dirac operator only acts on the left hand sector. It then becomes natural to define a local entropy.  The condition of local equilibrium then fully determines the distance function and, subsequently, the geometry.  When the temperature is independent of position, Euclidean geometry results with a scale determined by the global temperature.  Remarkably the scale increases as the temperature is lowered, leading to expanding distances at lower temperature.

There are a number of open issues worthwhile exploring further.  This includes the consequence of the stronger technical condition on the involutive algebra imposed here and the issue of the modified distance function that invokes the trace rather than the operator norm.  Another interesting point to explore further would be to investigate the class of Dirac operators for which (\ref{connesdq}) holds for all states and the implied associated geometries.
     

\noindent{\bf Acknowledgements.}
Support under the Indo-South African research agreement between the Department of Science and Technology, Government of India and the National Research Foundation of South Africa is acknowledged, as well as a grant from the National Research Foundation of South Africa.  



\begin{thebibliography}{99}
\bibitem{connes} A. Connes, Non-commutative Geometry, Academic Press, 1994. 
\bibitem{Doplicher} S.Doplicher, K.Fredenhagen and J.E.Roberts, Comm.Math.Phys. 172,187 (1995). 
\bibitem{Seiberg} N.Seiberg and E.Witten JHEP 09,032 (1999).
\bibitem{a6} A. Connes and M. Dubois-Violette, Comm. Math.Phys.230,539(2002); A.Connes and G.Landi, Comm. Math. Phys.221,141(2001).
\bibitem{a7} V. Gayral, J.M.Gracia-Bondia, B.Iochum, T.Schuker and J.C.Varilly, Comm. Math. Phys. 246,569(2004); V. Gayral and B. Iochum, J. Math Phys. 46,17(2005).
\bibitem{cag} E. Cagnache, F. D'Andrea, P Martinetti and J-C Wallet, Jnl. Geom. and Phys. 61, 1881 (2011).
\bibitem{mart} P. Martinetti and L Tomassini, arxiv:1110.6164 [math-ph].
\bibitem{a8} For recent reviews see K.Van den Dungen and W.D. van Suijlekom,"Particle Physics from Almost Commmutative Spacetimes", arxiv:1204.0328[hep-th]; M. Sakellaridou, "non-commutative spectral geometry: A guided tour for theoretical physicists", arXiv:1204.5772[hep-th].
\bibitem{a9} P. Basu, B.Chakraborty and F.G. Scholtz, J. Physics. A.44,285204(2011).
\bibitem{a10} F.G. Scholtz, B.Chakraborty, J Govaerts and S.Vaidya, J.Phys. A40,14581(2007).
\bibitem{a11} F.G.Scholtz, L.Gouba, A.Hafver, C.M.Rohwer J.Phys.A 42,175303 (2009).
\bibitem{a12} V. Galikova and P. Presnajder, arXiv:1112.4643[math-ph].
\bibitem{chris} C. M. Rohwer, "Additional degrees of freedom associated with position measurements in
  non-commutative quantum mechanics", arXiv:1206.1242 [hep-th]. 
\bibitem{a13} C. M. Rohwer, K.G. Zloshchastiev, L. Gouba  and F G Scholtz, J. Phys. A43, 345302 (2010).
\bibitem{a14} J.P.Provost and G. Vallee, Comm. Math. Phys.76,289(1980).
\bibitem{ash} A. Ashtekar and T.A. Schilling, "Geometrical Formulation of Quantum Mechanics"', arXiv:9706069 [gr-qc].
\bibitem{a15}  O. Bratelli and D.W. Robinson, Operator Algebras and Quantum Statistical Mechanics, Second edition, Vol 1., Springer, 1987, p.76. 
\bibitem{note1}It can be easily verified that the overlap $_{M}( \vec{x} |m,n )$ of the basis $|m,n)=|m\rangle\langle n|$ with the Moyal basis $|\vec{x} )_{M} = \dfrac{1}{2\pi} \int d^{2}p e^{-i\vec{p}\cdot \vec{x}}|\vec{p} )$, as introduced in \cite{a9}, 
indeed gives the Moyal representation $h_{mn}(\vec{x})$ used in \cite{cag,mart}.
 However, since the Moyal basis does not conform to the requirements of POVM \cite{a9}, it is desirable to carry out this analysis using the Voros basis or, better still, to carry this out in an abstract framework i.e. in a basis independent approach$-$as we do here.
\end{thebibliography}
\end{document}